\documentclass[aps,prd,preprint,nofootinbib]{revtex4-1}
\usepackage{amsfonts}
\usepackage{mathrsfs}
\usepackage{graphicx}
\usepackage{amsmath}
\usepackage{amssymb}
\usepackage{epsfig}
\usepackage{graphicx}
\usepackage{slashed}
\usepackage{geometry}
\usepackage[figuresleft]{rotating}

\usepackage{color}
\usepackage{multirow}
\usepackage{ulem}
\usepackage{adjustbox}

\usepackage{stmaryrd}
\newcommand{\bqa}{\begin{eqnarray}}
	\newcommand{\eqa}{\end{eqnarray}}
\newcommand{\beq}{\begin{equation}}
	\newcommand{\eeq}{\end{equation}}
\allowdisplaybreaks[1]

\graphicspath{{fig/}{dia/}} \DeclareGraphicsExtensions{.eps}
\begin{document}
	
	\title{Nonleptonic decays of $\Xi_{cc}\to \Xi_c \pi$ with $\Xi_c-\Xi_c'$ mixing}

	\author {Chia-Wei Liu\footnote{chiaweiliu@ucas.ac.cn} and Chao-Qiang Geng}
	%\email[Electronic address: ]{jinxiangnan21@mails.ucas.ac.cn}
	\affiliation{	
		School of Fundamental Physics and Mathematical Sciences, Hangzhou Institute for Advanced Study, UCAS, Hangzhou 310024, China\\
		University of Chinese Academy of Sciences, 100190 Beijing, China
	}
	\date{\today}

	\begin{abstract}
Aiming on testing the $\Xi_c-\Xi_c'$ mixing, we study the decays of $\Xi_{cc}\to \Xi_c \pi$ with $\Xi_{cc} = (\Xi_{cc}^{++} , \Xi_{cc} ^+ )$, $\Xi_c = (\Xi_c^{(\prime)+},\Xi_c^{(\prime)0})$ and $\pi = (\pi^+ , \pi^0)$. The soft-meson limit is considered along with the pole model, and the baryon matrix elements are evaluated by the bag model with and without removing the center-of-mass motion~(CMM). We  find that the four-quark operator matrix elements are about  twice larger once the unwanted CMM is removed.  We obtain that ${\cal R} = {\cal B}(\Xi_{cc}^+ \to \Xi_c^{\prime +} \pi^+ )/ {\cal B}(\Xi_{cc}^+ \to \Xi_c^{ +} \pi^+ ) = 0.87^{+0.17}_{-0.11} $ and $1.45$ with and without removing the CMM, where the former is close to the lower bound and the later is well consistent with  ${\cal R} = 1.41 \pm 0.17 \pm 0.10$ measured at LHCb. In addition, we show that after including the mixing,  the up-down asymmetry of  $\alpha( \Xi_{cc}^+ \to \Xi_c^{(\prime)0 } \pi^+)$ flips sign. Explicitly, we obtain  that $\alpha(\Xi_{cc}^{+} \to   \Xi_c^{\prime +} \pi^0) = 0.52$ and $\alpha(\Xi_{cc}^{+}  \to  \Xi_c^{ 0 } \pi^+) = 0.31$ with and without the CMM corrections, respectively, which are  all negative if the mixing is absence. As a bonus,  a positive value of  $\alpha(\Xi_{cc}^{+} \to \Xi_c^{\prime 0} \pi^+)$  in experiments  can also  serve as the evidence of the $W$-exchange contributions. 

	\end{abstract}

	\maketitle
	
\section{Introductions}

The baryon wave functions are the precondition in evaluating the decay quantities. It has been shown that the large $SU(3)$ flavor~($SU(3)_F$)  breaking effect in the singly charmed baryon semileptonic decays can be traced back to the $\Xi_c-\Xi_c'$ mixing~\cite{Geng:2022yxb,SU(3)1,Belle:2021crz,ALICE:2021bli}, given as
\begin{eqnarray}\label{ph}
	|\Xi_c  \rangle =  \cos\theta _c | \Xi_c^{ \overline{{\bf 3}}} \rangle +  \sin \theta_c  | \Xi_c^{  {\bf 6}}  \rangle  \,, \qquad
	|\Xi_c  ' \rangle =    \cos \theta_c | \Xi_c^ {{\bf 6}}  \rangle  -  \sin \theta_c | \Xi_c^{ \overline{{\bf 3}}}  \rangle 	\,,
\end{eqnarray}	 
where $\Xi_c^{(\prime)} = \Xi_c^{(\prime)+,0} $ are the physical baryons, and $\Xi_c^{ \overline{{\bf 3}}  ({\bf 6})}$ correspond to the antitriplet (sextet) charmed baryons. 
At the limit of the  $SU(3)_F$ symmetry, the physical baryons shall have definite $SU(3)_F$ representations, {\it i.e.} $\theta_c =0 $. From the mass relations, we have found that~\cite{Jenkins:1996rr}
\begin{equation}
\theta_c  = \pm  0.137 (5)\pi \,, 
\end{equation}
with the sign unfixed. 
In the decays involving $\Xi_c$, the mixing should be considered seriously as its effects are shown to be sizable~\cite{Geng:2022yxb}. 
It particular, it can be attributed to the nonzero signals of $\Xi_c^+ \to \Xi^{\prime 0}(1530) \pi^+$ 
observed at Belle~\cite{De0}, which are unexpected in the previous studies in the literature~\cite{De1,De2,De3,De4}.
 If the mixing is further confirmed,  it would undoubtedly reshape our knowledge of the baryon spin-flavor structures.

Recently, the LHCb collaboration has reported the ratio~\cite{Ratio}
\begin{equation}\label{expratio}
{\cal R}(\Xi_c^{++}\to \Xi_c^+ \pi^+)  = 1.41 \pm 0.17 \pm 0.10 \,,
\end{equation}
where 
${\cal R}(\Xi_{cc} \to \Xi_c \pi  ) \equiv{{\cal B}(\Xi_{cc} \to \Xi_c^{\prime } \pi )}/{{\cal B}(\Xi_{cc} \to \Xi_c \pi)}$, and 
the first and second uncertainties are systematic and statistical, respectively. 
It provides an ideal place to examine the mixing as it affects both the denominator and numerator of ${\cal R}$. In the literature~\cite{Cheng:2020wmk,Ke1,Gerasimov:2019jwp,Gutsche:2018msz,Wang:2017mqp,Sharma:2017txj} before the experiments, the ratio deviates largely to the value in Eq.~\eqref{expratio}.  In this work, we will show that the responsible mechanism is precisely the $\Xi_c-\Xi_c'$ mixing. 
On the other hand, combing several experiments, we have~\cite{LHCb:2018pcs,Belle:2019bgi,pdg}
\begin{equation}
\frac{{\cal B} ( \Xi_{cc} ^{++} \to \Xi_c^+ \pi ^+ ) }{{\cal B} ( \Xi_{cc} ^{++} \to \Lambda_c^+ K^- \pi^+  \pi ^+ )  }=  0.35  \pm 0.20  \,. 
\end{equation}	
By using  ${\cal B} ( \Xi_{cc} ^{++} \to \Lambda_c^+ K^- \pi^+  \pi ^+ ) 
>  {\cal B} ( \Xi_{cc} ^{++} \to \Sigma_{cc}^{++}  \overline{K}^{*0}  ) {\cal B}(\overline{K}^{*0} \to K^- \pi^+ )   $, 
${\cal B} ( \Xi_{cc} ^{++} \to \Sigma_{c}^{++}  \overline{K}^{*0}  )= 5.61\%$~\cite{Gutsche:2019iac} and 
${\cal B}(\overline{K}^{*0} \to K^- \pi^+) = 2/3$,  we obtain
\begin{equation}\label{lower}
{\cal B} ( \Xi_{cc} ^{++} \to \Xi_c^+ \pi ^+ )  > 0.59 \% \,,
\end{equation}
at $1\sigma$ confidence level. 
In addition, we have
\begin{equation} \label{expbr}
{\cal B} (\Xi_{cc}^{++} \to \Xi_c^+ \pi^+ ) = (1.33 \pm 0.74 ) \%\,,
\end{equation}
by assuming that the decay of $\Xi_{cc}^{++} \to \Lambda_c^+ K^- \pi^+ \pi^+$ contributes solely by $\Xi_{cc}^{++} \to \Sigma_{cc}^{++}  \overline{K}^{*0} $. 

On the theoretical aspect, 
it is known that a trustworthy method in evaluating the charm quark baryonic decays has not been given yet, since the charm quark is neither heavy nor light enough to apply the heavy quark  or  $SU(4)_F$ symmetry.  Nevertheless, it has been shown in Ref.~\cite{Lcthe} that the pole model conjunction with the current algebra and  soft-meson limit can well explain the experimental data of $\Lambda_c^+ \to B P$, with $B$ and $P$ the octet baryons and pseudoscalar mesons, respectively.  
As a phenomenological study, focusing on the mixing effects, we shall follow their methodology for the formalism. For the baryon wave functions,  we will examine both the mixing effects and the center-of-mass motion~(CMM) corrections of  the bag model. Very recently, it has been shown that the bag model is well consistent with the  experimental data of ${\cal B}(\Xi_Q  \to \Lambda_Q \pi ^-)$ once the CMM is removed~\cite{HFC,HFCBelle,HFCExp,HFCLHCb}, where 
$\Lambda_Q = (\Lambda_c^+, \Lambda_b^0)$ for
$\Xi_Q = (\Xi_c ^0 , \Xi_b ^- ) $. 

This work is organized as follows. In Sec.~\ref{2}, we briefly recall the formalism of the pole model and current algebra.  In Sec.~\ref{3}, we give the baryon wave functions  and their matrix elements with and without  the CMM. In Sec.~\ref{4}, we give the numerical results. We  conclude this study in Sec.~\ref{5}.

\section{formalism}	\label{2}

In general, the amplitude of ${\cal B}_i \to {\cal B}_f \pi$
 is decomposed as 
\begin{equation}\label{amplitudess}
 i \overline{u}_{f} (A - B\gamma _5 ) u_{i}\,,
\end{equation}
where $u_{i(f)}$ is the Dirac spinor of the initial~(final) baryon, and $A~(B)$ is referred to as the  parity violating~(conserving) amplitude. 
In the pole approximation, the nonfactorizable amplitudes read as~\cite{Cheng:2020wmk}
 \begin{equation}
\begin{aligned} 
&A^{\text {pole }}=-\sum_{\mathcal{B}_n^*}\left[\frac{g_{\mathcal{B}_f \mathcal{B}_n^* \pi } b_{n^* i}}{M_{i}-M_{n}^* }+\frac{b_{f n^*} g_{\mathcal{B}_n^* \mathcal{B}_i \pi }}{M_f-M_{n}^*}\right] \,, \\
&
B^{\text {pole }}=\sum_{\mathcal{B}_n}\left[\frac{g_{{\cal B}_i \mathcal{B}_n \pi } a_{n i}}{M_{i}-M_n}+\frac{a_{f n} g_{\mathcal{B}_n \mathcal{B}_i \pi }}{M_f-M_n}\right]\,,
\end{aligned}
 \end{equation}
where  ${\cal B}_{n}^{(*)}$ are the
parity even~(odd)
 intermediate baryons,  $M_{i,f,n}^{(*)}$  correspond to the  masses of ${\cal B}_{i,f,n}^{(*)}$, 
 \begin{equation}
\langle {\cal B} _2| {\cal H}_{eff} | {\cal B}_1  \rangle  = 
\overline{u}_2 \left( 
a_{2 1}+ b_{2 1} \gamma_5 
\right)  u_1 \,,~~~
\langle {\cal B} _n^*| {\cal H}_{eff} | {\cal B}_1  \rangle  = 
b_{n^* 1}\overline{u}_n 
  u_1 \,, 
 \end{equation} 
${\cal B}_{1,2} \in \{
{\cal B}_i, {\cal B}_f, {\cal B}_n
\}$, and 
${\cal H}_{eff}$ represents the effective Hamiltonian. 
The baryon-baryon-pion couplings of $g_{ {\cal B}_1{\cal B}_n^{(*)} \pi } $ are extracted by the
 Goldberg-Treiman relations
 \begin{equation}
g_{ {\cal B}_1{\cal B}_2 \pi }  = \frac{\sqrt{2}}{f_\pi} (M_1 + M_{2} )
g^{A(\pi)}_{{\cal B}_1{\cal B}_2} \,,~~~ 
g_{ {\cal B}_n^*{\cal B}_2 \pi }  = \frac{\sqrt{2}}{f_\pi} ( M_n^*- M_2 )
g^{A(\pi)}_{{\cal B}_n^*{\cal B}_2 } \,,
 \end{equation}
where 
 $f_\pi $ is the pion decay constant,
the axial vector couplings  of $g^{A(\pi)}_{{\cal B}' {\cal B}} $  are defined by 
\begin{equation}
 \langle {\cal B} ' | A^\mu(\pi)  | {\cal B}  \rangle = 
\overline{u'}  \left(
g^{A(\pi)}_{{\cal B}' {\cal B}} 
\gamma^\mu  - i \overline{g}_2 \sigma^{\mu \nu}q_{\nu} + \overline{g}_3 q^{\mu} 
\right)\gamma_5  u\,,
\end{equation} 
$u^{(\prime)}$ is the Dirac spinor of ${\cal B}^{(\prime)}$,  
$
A^\mu(\pi^+) = \overline{d} \gamma^\mu \gamma_5 u \,, ~ A^\mu(\pi^0) = \frac{1}{2}
\left(
\overline{u} \gamma^\mu \gamma_5 u - \overline{d} \gamma^\mu \gamma_5 d
\right),$
and ${\cal B}^{(\prime)} \in \{ {\cal B}_i , {\cal B}_f , {\cal B}_n, {\cal B}_n^* \}$.   Note that $\overline{g}_{2,3}$ are irrelevant  to this work.

To overcome the unknown baryon wave functions of ${\cal B}_n^*$, we use
 the soft-meson limit and $[Q_5^\pi +Q^\pi , {\cal H}_{eff}] = 0$\footnote{
The charge operators are defined as  $Q^{\pi}  = \int d^3 x ( q^\dagger \sigma_i q )/2 $  and 
$Q^\pi_5  = \int d^3 x ( q^\dagger \gamma_5  \sigma_i q )/2 $, where $q= (u,d)^T$ and $\sigma_i=\sigma_3, (  \sigma_1 \pm   i \sigma_2) / \sqrt{2} $ for $\pi = \pi^0, \pi^\pm$, respectively. 
The commutation relations come from  that the left-handed and right-handed currents commute. 
}.
The amplitudes of $\Xi_{cc} \to \Xi_c \pi$  are summarized   as\footnote{We note that the amplitudes of the charm baryon nonleptonic two-body decays~(196 in total) are compactly expressed by five of the topological tensor invariants within the current algebra~\cite{Groote:2021pxt}. }
~\cite{Cheng:2020wmk}
\begin{eqnarray}
A (\Xi_{cc}  ^{++} \to \Xi_c^{(\prime)+}\pi^+ ) &=&  \zeta\left( 
f_\pi^2 a_1 f_1^{(\prime) }  M_- ^{(\prime)}
- c_- a^{(\prime)} 
\right) \,,
\nonumber\\
A (\Xi_{cc}  ^{+} \to \Xi_c^{(\prime)0}\pi^+ ) &=&  \zeta \left( f_\pi^2  a_1f_1^{(\prime) } M_- ^{(\prime)}+  c_- a^{(\prime)}  \right) \,,\nonumber\\
A (\Xi_{cc}  ^{+} \to \Xi_c^{(\prime)+}\pi^0 ) &=&\sqrt{2}  \zeta  c_-   a^{(\prime)} \,,
\end{eqnarray}
and 
\begin{eqnarray}
&&B (\Xi_{cc}  ^{++} \to \Xi_c^{(\prime)+}\pi^+ ) =  \zeta\left( 
-	f_\pi^2 a_1g_1^{(\prime) } M_+^{(\prime)}
	-2 c_- a^{(\prime)} \frac{M_{cc}}{M_-^{(\prime)}} g^{A(\pi^+)}_{\Xi_{cc}^+ \Xi_{cc}^{++}}
	\right) \,, \\
&&B (\Xi_{cc}  ^{+} \to \Xi_c^{(\prime)0}\pi^+ ) =  \nonumber\\
&&~\zeta \left( -	f_\pi^2 a_1g_1^{(\prime) } M_+^{(\prime)} + c_- a \frac{M_{c} + M_c^{(\prime)}}{M_-} g^{A(\pi^+)}_{\Xi_{c}^{(\prime)0} \Xi_{c}^{ +} }
+c_- a' \frac{M'_{c} + M_c^{(\prime)}}{M_-'} g^{A(\pi^+)}_{\Xi_{c}^{(\prime)0}\Xi_{c}^{\prime +} }
\right)\,, \nonumber\\
&&B (\Xi_{cc}  ^{+} \to \Xi_c^{(\prime)+}\pi^0 ) = \nonumber\\
&&~\sqrt{2}\zeta c_-  \left(  -
2 a^{(\prime)} \frac{M_c}{M_-^{(\prime)}} g^{A(\pi^0)}_{\Xi_{cc}^+ \Xi_{cc}^+}
+ a \frac{M_c+M_c^{(\prime)}}{M_-} g^{A(\pi^0)}_{\Xi_{c}^{(\prime)+} \Xi_{c}^+}
+ a' \frac{M_c' + M_c^{(\prime)}}{M_-'} g^{A(\pi^0)}_{\Xi_{c}^{(\prime)+} \Xi_{c}^{'+}}
\right)\,, \nonumber
\end{eqnarray}
where
\begin{equation}
\zeta = \frac{	G_F}{f_\pi \sqrt{2}}V_{cs}V^*_{ud}\,,~~~c_- =\frac{1}{2}\left(
c_1 - c_2 
\right)\,,~~~M_\pm ^{(\prime)}= M_{cc} \pm  M_c^{(\prime)}\,,
\end{equation}
$M_{cc}$  and $M_c^{(\prime)}$ are the masses of $\Xi_{cc}$ and $\Xi_c^{(\prime)}$, respectively,
$G_F$ is the Fermi constant, 
$a_1$ is the effective Wilson coefficient,  and 
$V_{cs}$ and $V_{ud}$ are the Cabibbo-Kobayashi-Maskawa matrix elements. The information  of the baryon wave functions is encapsulated in $a$, $f_1$ and $g_1$, defined by\footnote{We use the Fierz transformation to sort $O_-$ defined in Ref.~\cite{Cheng:2020wmk}.}
\begin{eqnarray}
&&\langle \Xi_c ^{(\prime) +}  | 
O  | \Xi_{cc}^{+}\rangle  = \langle \Xi_c ^{(\prime) +}  | 
2( u^\dagger L^\mu d  ) 
(s^\dagger L_\mu c )  | \Xi_{cc}^{+}\rangle =  \overline{u}_c
\left(
a^{(\prime)}  + b^{(\prime)} \gamma_5 
\right)
 u_{cc}\,, \label{adefine}\\
&& \langle \Xi_c^{(\prime)+} | \overline{s}\gamma^\mu  c |\Xi_{cc} ^{++} \rangle = 
\overline{u}_c  \left(
f_1^{(\prime)} (\omega^{(\prime)}) \gamma^\mu  - i f^{(\prime)} _2(\omega^{(\prime)}) \frac{\sigma^{\mu \nu}}{M_{cc}}q_{\nu} + f^{(\prime)} _3(\omega^{(\prime)})\frac{q^{\mu}}{M_{cc}}
\right) u_{cc}\,, \nonumber \\
&& \langle \Xi_c^{(\prime)+} | \overline{s}\gamma^\mu \gamma_5 c |\Xi_{cc}^{++}  \rangle = 
\overline{u}_c  \left(
g_1^{(\prime)}(\omega^{(\prime)}) \gamma^\mu  - i g ^{(\prime)}_2(\omega^{(\prime)}) \frac{\sigma^{\mu \nu}}{M_{cc}}q_{\nu} + g^{(\prime)} _3(\omega^{(\prime)})\frac{q^{\mu}}{M_{cc}}
\right)\gamma_5  u_{cc}\,,\label{define}
\end{eqnarray}
with 
$L^\mu = \gamma^0 \gamma^\mu ( 1 -\gamma_5)$ and 
$u_{c(c)}$ the Dirac spinor of $\Xi_{c(c)}$. 
Since $\Xi_c^{\overline{{\bf 3}}} $
and $\Xi_c^{\bf 6}$ do not have  definite masses for $\theta_c \neq 0$, we define the variables 
\begin{equation}
	\omega^{(\prime)} = \frac{1+v^2}{1-v^2} = \frac{M_{cc}^2 + M_c^{(\prime)2} - M_\pi^2 }{2  M_c^{(\prime)} M_{cc} }\,,
\end{equation}
with $v$  the speed of the baryons in the Briet frame. 
Throughout this work, we employ the isospin symmetry, so that $\Xi_{cc} ^{++}~(\Xi_c^+)$ and $\Xi_{cc}^{+}~(\Xi_c^0)$ have the same masses and form factors.  
In addition, we have 
\begin{equation}
g_{\Xi_{cc}^+ \Xi_{cc}^{++}} ^{A(\pi^+)} =
- \frac{1}{2}
g_{\Xi_{cc}^+ \Xi_{cc}^{+}} ^{A(\pi^0)} \,,~~~g_{\Xi_{c}^{\prime 0} \Xi_{c}^{(\prime)+}} ^{A(\pi^+)} =
 \frac{1}{2}
g_{\Xi_{c}^{\prime +} \Xi_{c}^{(\prime)+}} ^{A(\pi^{0})}\,,~~~
g_{\Xi_{c}^{ 0} \Xi_{c}^{(\prime)+}} ^{A(\pi^+)} =
\frac{1}{2}
g_{\Xi_{c}^{ +} \Xi_{c}^{(\prime)+}} ^{A(\pi^{0})}\,.
\end{equation}
The above results are the  general ones under the soft-meson limit, and  the unknown parts of the baryon wave functions are absorbed in the form factors and $a^{(\prime)}$.  

Plugging the mixing of Eq.~\eqref{ph}  into  Eq.~\eqref{define}, we arrive at 
\begin{equation}
\begin{aligned}
&f_1 = \cos \theta_c  f_1^{\overline{{\bf 3}}} (\omega)+\sin \theta_c  f_1^{{\bf 6}}(\omega),~~~&g_1 = \cos \theta_c  g_1^{\overline{{\bf 3}}}(\omega) +\sin \theta_c  g_1^{{\bf 6}} (\omega)\,,~&\\
&f_1' = \cos \theta_c  f_1^{{\bf 6}}(\omega') -\sin \theta_c  f_1^{\overline{{\bf 3}}} (\omega'),~~~&g_1' = \cos \theta_c  g_1^{{\bf 6}} (\omega')-\sin \theta_c  g_1^{\overline{{\bf 3}}}(\omega')  \,,&\\
&a = \cos \theta_c a( {\overline{{\bf 3}}})+\sin \theta_c a({\bf 6})\,,~~~
&a '= \cos \theta_c a({\bf 6}) -  \sin \theta_c a(\overline{{\bf 3}}) \,,~~~~~&
\end{aligned}
\end{equation}
where $(f_1^{\overline{{\bf 3}}}(\omega^{(\prime)}) , f_1^{{\bf 6}}(\omega^{(\prime)})  $,  $(g_1^{\overline{{\bf 3}}}(\omega^{(\prime)}) , g_1^{{\bf 6}}(\omega^{(\prime)})  $
and $(a(\overline{{\bf 3}}) ,a({\bf 6})  )$ are calculated by taking $(
\Xi_c^{(\prime )} = \Xi_c(\overline{{\bf 3}} ) , \Xi_c^{(\prime )} =  \Xi_c({{\bf 6}} ) )$ in Eqs.~\eqref{adefine} and \eqref{define}. 
Similarly, the axial vector couplings are modified as 
\begin{eqnarray}
&&g_{\Xi_{c}^{\prime 0} \Xi_{c}^{\prime+}} ^{A(\pi^+)}  = 
\cos ^2 \theta_c g_{ {{\bf 6} } {\bf 6} } ^{A}  - \sin (2 \theta_c) g_{{\bf 6} \overline{{\bf 3} }}^A\,,\nonumber\\
&& g_{\Xi_{c}^{0} \Xi_{c}^{\prime+}} ^{A(\pi^+)}  =  g_{\Xi_{c}^{\prime 0} \Xi_{c}^{+}} ^{A(\pi^+)}  =
\cos (2 \theta_c) g_{ {{\bf 6} } \overline{{\bf 3}} } ^{A}  + \frac{1}{2}\sin (2 \theta_c) g_{{\bf 6} {\bf 6 }}^A\,,\nonumber\\
&&g_{\Xi_{c}^{0} \Xi_{c}^{+}} ^{A(\pi^+)}  = 
\sin (2 \theta_c ) g_{{\bf 6} \overline{{\bf 3} }}^A  +  \sin^2 \theta_c g_{ {{\bf 6} } {\bf 6} } ^{A} \,,
\end{eqnarray}
with 
\begin{equation}
\langle \Xi_c^0 ({\bf R}_2) | \overline{d} \gamma^\mu \gamma_5 u   | \Xi_c^+ ({\bf R}_1)  \rangle = 
\overline{u}_{{\bf R}_2} \left(
g^{A}_{{\bf R}_2{\bf R}_1} 
\gamma^\mu  - i \overline{g}_2 {\sigma^{\mu \nu}}q_{\nu} + \overline{g} _3 {q^{\mu}}
\right)\gamma_5  u_{{\bf R}_1}
\end{equation}
and ${\bf R}_{1,2} = (\overline{{\bf 3}}, {\bf 6} )$.
Finally, the decay widths and up-down asymmetries are  given by
\begin{equation}
\begin{aligned}
&\Gamma=\frac{{\bf p}_f}{8 \pi}\frac{ (M_i + M_f) ^{ 2}-M_\pi^2}{M_{i}^2}\left( |A|^2+ \kappa ^2 |B|^2\right)\,,\\
&\alpha=\frac{2 \kappa \operatorname{Re}\left(A^* B\right)}{|A|^2+\kappa^2|B|^2}\,,
\end{aligned}
\end{equation}
where ${\bf p}_f$ is the magnitude of the pion  three-momentum, and $\kappa= {\bf p}_f/( E_f + M_f)$ with $E_f = \sqrt{{\bf p}_f^2 + M_f^{2}} $.

\section{Baryon wave functions and matrix elements}\label{3}

The bag model provides  approximations of the hadron wave functions, aiming on reconciling two very different ideas in QCD~\cite{Bag1,Bag2,Bag3}. Inside the bag, quarks move freely as a result of  the asymptotic freedom but can not penetrate the bag due to the QCD confinement. 
One of the great advantages of the bag model is that the parameters are fitted from the mass spectra. 
Consequently, the model provides fixed predicted results, which can be tested by the experiments.
%Though the model contains uncontrollable systematic errors, it provides fixed results for the experiments to testify.}
In this work, we calculate the baryon matrix elements by the bag models with and without removing the CMM, referred to as the homogeneous bag~(HB) and static bag~(SB) approaches, respectively.  

The baryon wave functions concerned by this work are given as 
\begin{eqnarray}
&&	|\Xi _{cc}, \updownarrow\rangle = \int\frac{1}{2\sqrt{3} } \epsilon^{\alpha \beta \gamma} q _{a\alpha}^{\dagger} (\vec{x}_1) c_{b\beta}^\dagger(\vec{x}_2) c_{c\gamma}^\dagger (\vec{x}_3) \Psi_{A_\updownarrow(ucc)}^{abc} (\vec{x}_1,\vec{x}_2,\vec{x}_3) [d^3  \vec{x}] | 0\rangle\,,\nonumber\\
&&	|\Xi _{c}^{\overline{{\bf 3}}}, \updownarrow\rangle = \int\frac{1}{\sqrt{6} } \epsilon^{\alpha \beta \gamma} q _{a\alpha}^{\dagger} (\vec{x}_1) s_{b\beta}^\dagger(\vec{x}_2) c_{c\gamma}^\dagger (\vec{x}_3) \Psi_{A_\updownarrow(qsc)}^{abc} (\vec{x}_1,\vec{x}_2,\vec{x}_3) [d^3  \vec{x}] | 0\rangle\,,\nonumber\\
&&	|\Xi _{c}^{{\bf 6}}, \updownarrow\rangle = \int\frac{1}{\sqrt{6} } \epsilon^{\alpha \beta \gamma} q _{a\alpha}^{\dagger} (\vec{x}_1) s_{b\beta}^\dagger(\vec{x}_2) c_{c\gamma}^\dagger (\vec{x}_3) \Psi_{S_\updownarrow(qsc)}^{abc} (\vec{x}_1,\vec{x}_2,\vec{x}_3) [d^3  \vec{x}] | 0\rangle\,,
\end{eqnarray}
where $q_{a\alpha} \in \{ u_{a\alpha},d_{a\alpha} \}$, 
the Latin~(Greek) letters are the color~(Dirac spinor) indices, and $\Psi$ describe the spatial distributions of the quarks. In the SB, $\Psi$  read as~\cite{Bag3} 
\begin{eqnarray}\label{distri}
&&\Psi_{A \updownarrow (q_1q_2q_3)} ^{abc(\text{SB})} (\vec{x}_1,\vec{x}_2,\vec{x}_3)= 
\frac{{\cal N}}{\sqrt{2}}\Big( 
\phi^a_{q_1\uparrow}(\vec{x}_1)\phi^b_{q_2\downarrow}(\vec{x}_2)- \phi^a_{q_1\downarrow}(\vec{x}_1)\phi^b_{q_2\uparrow}(\vec{x}_2)
\Big) \phi^c_{q_3\updownarrow}(\vec{x}_3)
\,,\nonumber\\
&&\Psi_{S \uparrow (q_1q_2q_3)} ^{abc(\text{SB})} (\vec{x}_1,\vec{x}_2,\vec{x}_3)= 
\frac{{\cal N}}{\sqrt{6}}\Big( 
2 \phi^a_{q_1\uparrow}(\vec{x}_1)\phi^b_{q_2\uparrow}(\vec{x}_2) \phi^c_{q_3\downarrow}(\vec{x}_3)\nonumber\\
&&\quad\qquad- \phi^a_{q_1\downarrow}(\vec{x}_1)\phi^b_{q_2\uparrow}(\vec{x}_2) \phi^c_{q_3\uparrow}(\vec{x}_3)- \phi^a_{q_1\uparrow}(\vec{x}_1)\phi^b_{q_2\downarrow}(\vec{x}_2) \phi^c_{q_3\uparrow}(\vec{x}_3)
\Big)
\,,
\end{eqnarray}
where ${\cal N}$ is the normalization constant, 
\begin{equation}
\phi_{q\updownarrow}(\vec{x}) = \left(
\begin{array}{c}
\omega_{q_+}j_0({\bf p}_q|\vec{x}|) \chi _ \updownarrow\\
i\omega_{q_-} j_1 ({\bf p}_q|\vec{x}|) \hat{x} \cdot \vec{\sigma} \chi _ \updownarrow
\end{array}
\right)~~~~\text{for} |\vec{x}| < R,
\end{equation}
$R$ is the bag radius, 
 ${\bf p}_q$  is the magnitude  of the quark three-momentum, $\omega_{q_\pm} = \sqrt{E_q \pm M_q}$ with $M_q$ the quark mass and $E_q= \sqrt{{\bf p}_q^2 + M_q ^2}$, $j_{0,1}$ are the spherical Bessel functions,  $\chi_\uparrow= (1,0)^T$   and $\chi_\downarrow= (0,1)^T$.

The baryon wave functions in Eq.~\eqref{distri}  are localized and can not be momentum eigenstates according to the Heisenberg principle. 
 In other words, the baryons at rest must be invariant under the spatial translations so can not be localized.
Another way to see the  problem is that the spatial wave functions~(three-momenta)  of the quarks   in the SB are untangled. Therefore, we  have $\langle ({\bf p}_1+{\bf p}_2+ {\bf p}_3) ^2 \rangle = 
\langle {\bf p}_1^2\rangle +\langle {\bf p}_2^2\rangle + \langle {\bf p}_3^2\rangle
>0$, where ${\bf p}_{i}$ is the three-momentum of the $i$-th quark
%, and we have used 
with $\langle {\bf p}_{i} \rangle = 0 $ and $\langle {\bf p}_{i} {\bf p}_{j} \rangle = \langle {\bf p}_{i} \rangle \langle {\bf p}_{j} \rangle $ 
for $i\neq j$. To overcome the problem, the baryon wave functions shall be distributed uniformly over the three-dimensional space, 
while the quarks shall be entangled in the spatial wave functions. The simplest way to do the job is to linearly superpose the wave functions over the three-dimensional space~\cite{Liu:2022pdk}
\begin{equation}\label{26}
\Psi^{\text{(HB)}} (\vec{x}_1, \vec{x}_2 ,\vec{x}_3) = \int d^3 \vec{x}_\Delta \Psi^{\text{(SB)}}  (\vec{x}_1 - \vec{x}_\Delta , \vec{x}_2- \vec{x}_\Delta  ,\vec{x}_3- \vec{x}_\Delta ) \,,
\end{equation}
where $\Psi^{(\text{SB})}$  are the ones given in Eq.~\eqref{distri}. With this trick, the translational invariance
of the baryons is recovered since 
\begin{equation}\label{27}
\Psi^{\text{(HB)}} (\vec{x}_1+\vec{d}, \vec{x}_2 +\vec{d},\vec{x}_3+\vec{d} )=\Psi^{\text{(HB)}} (\vec{x}_1, \vec{x}_2 ,\vec{x}_3) \,,
\end{equation}
where $\vec{d}$ is an arbitrary three-vector, and the equality can be proved by taking $\vec{x}_\Delta \to \vec{x}_\Delta + \vec{d}$ in Eq.~\eqref{26}.  
From Eq.~\eqref{27}, it is clear that the quarks are no longer constrained in the specific region.
However, the quarks are bounded and entangled in the sense that 
\begin{equation}
\Psi^{\text{(HB)}} (\vec{x}_1, \vec{x}_2 ,\vec{x}_3) = 0\,,~~~\text{for}~~|\vec{x}_i - \vec{x}_j| >2R\,,
\end{equation}
for $i,j \in\{ 1,2,3\}$, which can be derived by $\phi_{q\updownarrow} (\vec{x}) = 0$ for $\vec{x}>R$.  As the baryons are invariant under the spatial translations, we conclude that the CMM is removed.

With the baryon wave functions, the calculations of the baryon matrix elements are straightforward.  The results of the SB approach can be found in Ref.~\cite{Cheng:2020wmk},  while  the form factors of the HB approach are given in Ref.~\cite{Geng:2022uyy}.

Here, we sketch the method of calculating 
 $a(\overline{{\bf 3}})$ and  $a({\bf 6})$ in the HB approach.  To diminish the directional  dependencies in Eq.~\eqref{adefine}, we trace over  the baryon spins
 \begin{equation}\label{19}
a({\bf R})= \frac{1}{2} \left( \langle \Xi_c ^{ +} ({\bf R}), \uparrow| O| \Xi_{cc}^+, \uparrow \rangle + 
\langle \Xi_c ^{ +}({\bf R}),\downarrow | O| \Xi_{cc}^+ ,\downarrow \rangle 
\right) \,.
 \end{equation}
with the normalization of  $\overline{u}_{c(c)} u_{c(c)} = 1$.  By
using the anticommutation  relations among the quark operators
\begin{equation}
\left\{q_{a \alpha}(\vec{x}), q_{b \beta}^{\dagger}\left(\vec{x}^{\prime}\right)\right\}=\delta_{a b} \delta_{\alpha \beta} \delta^3\left(\vec{x}-\vec{x}^{\prime}\right)\,,
\end{equation}
we arrive at~\cite{HFC} 
\begin{equation}\label{21}
\sum _{J_z=\updownarrow} 
\langle \Xi_c^{ + }({\bf R}) ,J_z | (u^\dagger L^\mu d) ( s^\dagger L_\mu c) | \Xi_{cc}^{+ }, J_z \rangle  = {\cal N}_c {\cal N}_{cc} 
\int d^3 \vec{x}_\Delta {\cal D}_c(\vec{x}_\Delta ) { \Upsilon}^{{\bf R}} (\vec{x}_\Delta )\,,
\end{equation}
where ${\cal N}_{c(c)}$ is the normalization constant of $\Xi_{c(c)}$,
\begin{eqnarray}\label{22}
&&{\cal D}_c (\vec{x}_\Delta ) = \int d\vec{x} \phi^\dagger_c \left( \vec{x}^+ \right) 
\phi_c \left( \vec{x} ^-  \right), \\
&&\Upsilon^{{\bf R}} (\vec{x}_\Delta) = 
\sum_{[\lambda]} {\cal F}([\lambda], {\bf R})
\int d^3 \vec{x} \phi_{u \lambda_4}^{\dagger}\left(\vec{x}^{+}\right) L_\mu \phi_{d \lambda_2}\left(\vec{x}^{-}\right) \phi_{s \lambda_3}^{\dagger}\left(\vec{x}^{+}\right) L^\mu \phi_{c \lambda_1}\left(\vec{x}^{-}\right) ,\nonumber
\end{eqnarray}
 $[\lambda] = (\lambda_1,\lambda_2,\lambda_3,\lambda_4)$,  $ \vec{x}^\pm = \vec{x} \pm \vec{x}_\Delta /2 $, and ${\cal F} $ are the spin-flavor overlappings, given as 
\begin{eqnarray}\label{spinflavor}
&&\sum_{[\lambda]}{\cal F}([\lambda],\overline{\bf 3}) \left( \lambda_1 \otimes \lambda_2 \otimes\lambda_3 \otimes \lambda_4\right) =  \frac{\sqrt{6} }{2} \left( 
\uparrow\downarrow\uparrow\downarrow - \downarrow\uparrow\uparrow \downarrow -\uparrow\downarrow\downarrow\uparrow+ \downarrow\uparrow\downarrow\uparrow
\right) \,, \\ 
&&\sum_{[\lambda]}{\cal F}([\lambda],{\bf 6}) \left( \lambda_1 \otimes \lambda_2 \otimes\lambda_3 \otimes \lambda_4\right) =  \frac{1 }{3\sqrt{2}} \left[
\left( \uparrow\downarrow + \downarrow \uparrow\right)  
\left(
\uparrow\downarrow + \downarrow \uparrow
\right) + 2 \uparrow\uparrow\uparrow\uparrow + 2 \downarrow \downarrow \downarrow \downarrow
\right]\,.\nonumber
\end{eqnarray}
From Eq.~\eqref{spinflavor}, it is easy to deduce that 
\begin{eqnarray}\label{overlapF}
	&&\sum_{[\lambda]} 
	{\cal F}([\lambda] , {\bf R})
	\left(\chi_{\lambda_3}^{\dagger} \chi_{\lambda_1}\right)\left(\chi_{\lambda_4}^{\dagger} \chi_{\lambda_2}\right)={\cal C}_{\text{unflip}}^{{\bf R}} \,,\nonumber\\
&&\sum_{[\lambda]} 
{\cal F}([\lambda] , {\bf R})
\left(\chi_{\lambda_3}^{\dagger} \sigma_i \chi_{\lambda_1}\right)\left(\chi_{\lambda_4}^{\dagger} \chi_{\lambda_2}\right)= 0\,,\nonumber\\
	&&\sum_{[\lambda]}  {\cal F}([\lambda] , {\bf R}) \left(\chi_{\lambda_3}^{\dagger} \sigma_i \chi_{\lambda_1}\right)\left(\chi_{\lambda_4}^{\dagger} \sigma_j \chi_{\lambda_2}\right)=\delta_{i j} {\cal C}_{\text{flip}}^{{\bf R}}  \,,
\end{eqnarray}
where
\begin{equation}
	\begin{aligned}
		\left({\cal C}_{\text{unflip} } ^{\overline{{\bf 3}}},{\cal C}_{\text{flip} } ^{\overline{{\bf 3}}}\right) = \left(\sqrt{6} , -\sqrt{6} \right) \,,~~~
		\left({\cal C}_{\text{unflip} } ^{{\bf 6}}, {\cal C}_{\text{flip} } ^{{\bf 6}}\right)  = \left(\sqrt{2},\frac{\sqrt{2}}{3 } \right)\,,
	\end{aligned}
\end{equation}
and  $\sigma_{i,j}$ are the Pauli matrices.  The second and third equations of Eq.~\eqref{overlapF}  are due to that 
we have traced over the baryon spins so the matrix elements can not depend on  specific directions.

We decompose $\Upsilon$ into several pieces
	\begin{equation}\label{col1}
\Upsilon^{\bf R}(\vec{x}_\Delta)  = \int d^3\vec{x} \sum_{k=1,2,3,4} \Gamma^{\bf R}_k(\vec{x}_\Delta,\vec{x}) 
\end{equation}
with 
\begin{eqnarray}\label{B3}
	\Gamma_1^{\bf R}(\vec{x}_\Delta,\vec{x})  &=&
	\sum_{[\lambda]}  {\cal F}([\lambda] , {\bf R})
	\phi _{u\lambda_4} ^\dagger\left(\vec{x}^ + \right) \phi_{d\lambda_2}\left(\vec{x}^- \right) \phi _{s\lambda_3} ^\dagger\left(\vec{x} ^+ \right) \phi_{c\lambda_1}\left(\vec{x} ^- \right) \,, \nonumber\\
	\Gamma_2^{\bf R}(\vec{x}_\Delta,\vec{x})&=&
	\sum_{[\lambda]}  {\cal F}([\lambda] , {\bf R})
	\phi _{u\lambda_4} ^\dagger\left(\vec{x}^ + \right) \gamma_5 \phi_{d\lambda_2}\left(\vec{x}^- \right) \phi _{s\lambda_3} ^\dagger\left(\vec{x} ^+ \right) \gamma_5 \phi_{c\lambda_1}\left(\vec{x} ^- \right)  \,,\nonumber\\
	\Gamma_3^{\bf R}(\vec{x}_\Delta,\vec{x})&=&-
	\sum_{[\lambda]}   {\cal F}([\lambda] , {\bf R})
	\phi _{u\lambda_4} ^\dagger\left(\vec{x}^ + \right) V_i \phi_{d\lambda_2}\left(\vec{x}^- \right) \phi _{s\lambda_3} ^\dagger\left(\vec{x} ^+ \right) V_i \phi_{c\lambda_1}\left(\vec{x} ^- \right) \,, \\
	\Gamma_4^{\bf R}(\vec{x}_\Delta,\vec{x})&=& -
	\sum_{[\lambda]}   {\cal F}([\lambda] , {\bf R})
	\phi _{u\lambda_4} ^\dagger\left(\vec{x}^ + \right) V_i\gamma_5 \phi_{d\lambda_2}\left(\vec{x}^- \right) \phi _{s\lambda_3} ^\dagger\left(\vec{x} ^+ \right) V_i \gamma_5\phi_{c\lambda_1}\left(\vec{x} ^- \right)  , \nonumber
\end{eqnarray}
where $V _i = \gamma_0 \gamma_i$ with $i=1,2,3$. Plugging  Eq.~\eqref{overlapF} into Eq.~\eqref{B3}, we obtain 
	\begin{eqnarray}\label{col2}
	\Gamma_1^{\bf R}(\vec{x}_\Delta,\vec{x}) &=&
{\cal C}_{\text{unflip}}^{{\bf R}} 	\left( 
	u_u^+ u_d^- +  v_u^+v_d^-  
	\hat{x}^+ \cdot \hat{x}^-  
	\right)\left( 
	u_s^+ u_c^- +  v_s^+v_c^-  
	\hat{x}^+ \cdot \hat{x}^-   
	\right) \nonumber\\
	&&-{\cal C}_{\text{flip}}^{{\bf R}} \frac{(\vec{x}_\Delta \times \vec{x}  ) ^2}{(r^+r^-)^2}  v_u^+v_d^-   v_s^+v_c^- \,, \nonumber\\
	\Gamma_2^{\bf R} (\vec{x}_\Delta,\vec{x}) &=&-{\cal C}_{\text{flip}}^{{\bf R}}   \left(   u_u^+v_d^- \hat{x}^- -  v_u^+  u_d^-  \hat{x}^+ \right) 
	\left( u_s^+  v_c^-  \hat{x}^- -  v_s^+  u_c^-  \hat{x}^+ \right) \,,\nonumber\\
	\Gamma_3^{\bf R}  (\vec{x}_\Delta,\vec{x})&=&  -\frac{ {\cal C}_{\text{unflip}}^{\bf R} } { {\cal C}^{\bf R}_{\text{flip}}}\Gamma_2^{\bf R} (\vec{x}_\Delta,\vec{x})
	-2 {\cal C}_{\text{flip}}^{\bf R}  \left( u_u^+  v_d^-  \hat{x}^- +  v_u^+  u_d^-  \hat{x}^+ \right) \cdot
	\left( u_s^+  v_c^-  \hat{x}^- +  v_s^+  u_c^-  \hat{x}^+ \right)  \,,\nonumber\\
	\Gamma_4^{\bf R}(\vec{x}_\Delta,\vec{x}) &=&  -{\cal C}_{\text{flip}}^{\bf R}   \Big[ 3 u_u^+u_d^- u_s^+ u_c^- +
	v_u^+v_d^- v_s^+ v_c^-  \left( 2 +(\hat{x}^+ \cdot\hat{x}^-)^2 \right)  \\
	&&-(u_u^+u_d^- v_s^+ v_c^- + v_u^+v_d^- u_s^+ u_c^-  ) \hat{x}^+\cdot \hat{x}^-\Big]+	 {\cal C}_{\text{unflip}}^{\bf R}  v_u^+ v_d^- v_s^+ v_c^-  \frac{   (\vec{x}_\Delta \times \vec{x}) ^2 }{(r^+r^-)^2} \,,\nonumber
\end{eqnarray}
with  the abbreviation
\begin{equation}
\phi_q \left(\vec{x} ^\pm \right) = \left(
\begin{array}{c}
u^\pm _q \chi \\ 
iv^\pm _q \left( \hat{x}^\pm \cdot \vec{\sigma} \right)  \chi 
\end{array}
\right)\,.
\end{equation}

Collecting Eqs.~\eqref{19}, \eqref{21}, \eqref{22}, \eqref{B3} and \eqref{col2},  now we are able to calculate $a({\bf R})$.  Note that the formalism is reduced to the SB approach  by eliminating the $\vec{x}_\Delta$ integral
\begin{equation}
a({\bf R}) =   \Upsilon^{\bf R} (0) \,.
\end{equation}
To compare with the SB approach~\cite{Cheng:2020wmk}, we rescale the parameters as 
\begin{equation}
	a(\overline{\bf 3})=16\sqrt{6} \pi  X_2, \quad a({\bf 6})=-\frac{16 \sqrt{2} \pi }{3} X_1 \,.
\end{equation}

\section{Numerical results }
\label{4}
In crunching up the numbers, we take the bag model parameters~\cite{bag}
\begin{equation}
	M_{u,d} = 0\,,~~~M_s = 0.28~\text{GeV}\,,~~~ M_c = 1.655~\text{GeV}\,,~~~R= (5.0\pm 0.1)~\text{GeV}^{-1}\,. 
\end{equation} 
In the HB model,
the axial vector couplings and $X_{1,2}$ are found to be 
\begin{equation}
\begin{aligned}
&\left(g^{A(\pi^+)}_{\Xi_{cc}^+\Xi_{cc}^{++}}\,,~ g^A_{{\bf 6}{\bf 6}}\,,~g^A_{ {\bf 6 }\overline{\bf 3}}\right) = (-0.259, 0.522, -0.453 )\,, \\
&X_2 = (3.52 \pm 0.22)  10^{-4}~\text{GeV}^3\,,~~~X_1 =  (-2.44\pm 0.08)  10^{-6}~\text{GeV}^3\,.
\end{aligned}
\end{equation}
while $g_1$ and $f_1$  are summarized in TABLE~\ref{tab: Overlaps}. 
The  overlappings of $X_2$ and $X_1$  are twice and one half larger than those of  the SB approach~\cite{Cheng:2020wmk}, and  the same tendencies are found in the heavy-flavor-conserving decays~\cite{HFC}.
We emphasize that $X_1 \propto M_s$ due to the K\"orner-Pati-Woo theorem~\cite{Korner:1970xq}.
 As a consequence, the calculated $X_1$ from the bag model shall not be fully trusted as $M_s$  is difficult to be determined. 
Nevertheless, $X_1$ can be taken as zero in practice  so the final results are little affected.

\begin{table}[t]
	\caption{\label{tab: Overlaps} Results of the form factors in the HB approach\footnote{
The uncertainties are smaller than the ones obtained in Ref.~\cite{Geng:2022uyy} for that a smaller range of the bag radii is considered. 
	}.}
	\centering
	\begin{tabular}{ccccccccccc}
		\hline\hline
		$f_1^{\overline{\bf 3}} (\omega)$ &$f_1^{\overline{\bf 3}} (\omega')$&$f_1^{{\bf 6}} (\omega)$ &$f_1^{{\bf 6}} (\omega')$&$g_1^{\overline{\bf 3}} (\omega)$ &$g_1^{\overline{\bf 3}} (\omega')$&$g_1^{{\bf 6}} (\omega)$ &$g_1^{{\bf 6}} (\omega')$ \\
		\hline
		$0.480(17)$ ~& $0.593(17)$~& $0.277(10)$~ & $0.342(10)$~&$0.152(5)$ ~& $0.188(5)$~&$0.439(16)$~ &$0.542(15)$~ \\ 
		\hline\hline
	\end{tabular}
\end{table}

\begin{figure}
	\includegraphics[width=0.6 \linewidth]{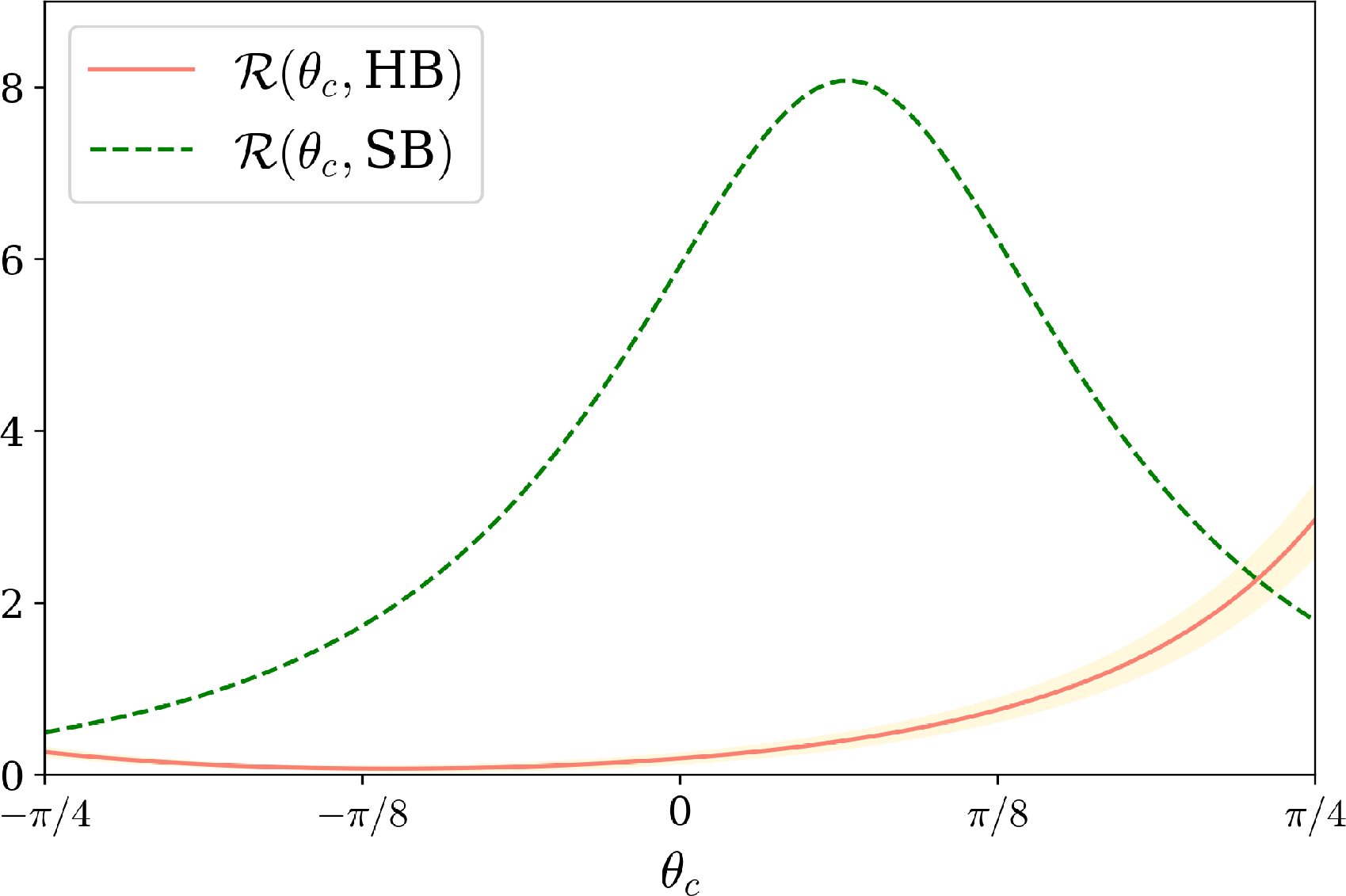} 
\caption{${\cal R}(\Xi_c^{++} \to \Xi_c^+\pi^+)$ versus $\theta_c$.}
\label{figure}
\end{figure}

The mixing largely modifies ${\cal R}(\Xi_c^{++} \to \Xi_c^+\pi^+)$, as shown in FIG.~\ref{figure}. Particularly, with $\theta_0 \equiv 0.142\pi$  we find  that 
\begin{equation}\label{Rtheo}
	\begin{aligned}
&{\cal R} (0 ,\text{SB} ) = 6.74\,,
&{\cal R} ( \theta_0 ,\text{SB} ) = 5.39\,, \qquad\quad\,&\quad {\cal R} ( -\theta_0 ,\text{SB} ) = 1.45\,,\\ 
&{\cal R} (  0 ,\text{HB} ) = 0.19\pm 0.05\,,
&{\cal R} ( \theta_0 ,\text{HB} ) = 0.87^{+0.17}_{-0.11} \,,~~ &\quad{\cal R} ( -\theta_0 ,\text{HB} ) = 0.07\,.
	\end{aligned}
\end{equation}
Due to the large difference in $X_{1,2}$, the HB and SB approaches predict  very different ratios.
However, they both require  $\theta_c \neq 0$ to explain the experiments. With $\theta_c = -\theta_0$, the SB approach is in good agreement with Eq.~\eqref{expratio}, whereas with $\theta_c = \theta_0$ the HB approach show accordance with the experimental lower bound.

We list out the results of the branching fractions and up-down asymmetries in TABLE~\ref{brresult} along with those in  the  literature, where we have normalized the branching fractions by $(\tau (\Xi_{cc}^{++}) , \tau(\Xi_{cc}^{+} ))= (2.56, 0.45) \times 10^{-13}s$~\cite{LifeTH,EXPLifetime}. 
In the literature,
Ref.~\cite{Gutsche:2018msz} adopts the covariant quark model  up to three-loop calculations, Ref.~\cite{Sharma:2017txj}  employs the pole model but only the parity even pole is considered, Ref.~\cite{Shi:2022kfa} calculates the $W$-exchange  contributions by the light cone sum rule with the  heavy quark effective theory, and Refs.~\cite{Gerasimov:2019jwp,Ke1,Ke2} consider  only the factorizable parts of the amplitudes. In the table, the quoted values of Ref.~\cite{Sharma:2017txj}  are calculated by the nonrelativistic quark model~(N) and heavy quark effective theory~(H) with the flavor-independent pole, and the ones of Refs.~\cite{Ke1,Ke2} are given by $\theta_c =
0.090 \pm 0.013 \pi$ 
 (M) and $\theta_c=0$~(N) with the light-front quark model. 
The  results of Ref.~\cite{Cheng:2020wmk} are essentially the ones of the SB approach with $\theta_c = 0$. 
Remarkably,   Refs.~\cite{Cheng:2020wmk} and \cite{Gutsche:2018msz} show a good accordance, which indicates their treatments 
for $\theta_c = 0$
are reliable. However, they are inconsistent
 with the experimental data
 of   ${\cal R}(\Xi_c^{++} \to \Xi_c^+\pi^+)$. 
We believe that such deviations are caused by  the $\Xi_c - \Xi_c'$ mixing. As shown in the table, after considering the mixing, both ${\cal B}$ and ${\cal R}$ are compatible with the current experimental data. 
To test our theory, we recommend the future experiments on ${\cal R}(\Xi_c^+\to \Xi_c^{0(+)} \pi ^{+(0)})$, found to be 
\begin{eqnarray}
&&{\cal R}(\Xi_c^+\to \Xi_c^{0} \pi ^{+}) = 0.25~(\text{SB}),~~1.17~(\text{HB})\,,\nonumber\\
&&{\cal R}(\Xi_c^+\to \Xi_c^{+} \pi ^{0}) = 0.23~(\text{SB}),~~0.25~(\text{HB})\,.
\end{eqnarray}
It is interesting to point out that the sign of $\alpha(\Xi_{cc}^+ \to \Xi_c^0 \pi^+)$ is flipped by the mixing in the SB approach. Under the factorization ansatz, the decays of $\Xi_{cc}^{++} \to \Xi_c^{(\prime) +} \pi^+$ and $\Xi_{cc}^{+} \to \Xi_c^0 \pi^+$ behave identically, {\it i.e.} they have the same decay widths and up-down asymmetries as shown explicitly in Refs.~\cite{Gerasimov:2019jwp,Ke1,Ke2}.  Therefore, the experimental measurements of ${\cal B}$ and $\alpha$ up on these decays may clarify the  $W$-exchange contributions. 
Especially, we  recommend the future measurements on $\alpha(\Xi_{cc}^{+} \to \Xi_c^{(\prime) 0} \pi^+)$ as it is essentially  negative in the factorization ansatz with $\theta_c=0$. It is interesting to point out that the sign of $\alpha(\Xi_{cc}^0 \to \Xi_c^0 \pi^+)$ is flipped after the mixing is considered in both the SB approach and Ref.~\cite{Ke2}.

Unfortunately, with the experimental value in Eq.~\eqref{expratio}, 
 the HB and SB models suggest opposite signs of $\theta_c$ as shown in Eq.~\eqref{Rtheo}.  
In  $\Xi_c^0 \to \Lambda_c^+ \pi^-$ and $\Xi_b^-\to \Lambda_b^0 \pi^-$, where the soft-meson limit is trustworthy\footnote{
The ${\bf p}_f$ in $\Xi_{cc}^{++} \to \Xi_c^+ \pi^+$ 
and $\Xi_{c}^{0} \to \Lambda_c^+ \pi^-$ are $0.96$ and $0.11$~GeV, respectively.  
}, it has been found that the HB approach is much  more suitable than the SB one~\cite{HFC,HFCBelle,HFCExp,HFCLHCb}. More importantly, the HB wave functions are self-consistent on the contrary of  the SB ones. 
However, the computed ${\cal B} (\Xi_{cc}^{++} \to \Xi_c^+ \pi^+)$  with   the HB   is  much larger than  Eq.~\eqref{expbr},   indicating  that the  branching fractions might be overestimated. Viewing  on  the successes of the SB approach in the $\Lambda_c^+$ decays~\cite{Lcthe}, it is  likely that the CMM and finite ${\bf p}_f$ corrections compensate each others. 
Accordingly, the sign of $\theta_c$ shall be negative, suggested by the SB model. 
We note that  the semileptonic decays of $\Xi_{cc} \to \Xi_c e^+ \nu_e$  are  ideal places to determine the sign of $\theta_c$, as they are uncontaminated  by the $W$-exchange contributions.  Nonetheless, the experiments are subjected to the difficulties imposed by the chargeless neutrinos. 
%A future  study on the charmed  baryon lifetimes, involving the four-quark operator matrix elements, will be able to settle down the issue. 

%Thus, measuring these quantities can test the validity of the factorization ansatz and the $\Xi_c -\Xi_c'$ mixing. 

\clearpage
\newgeometry{top=10mm} 
\begin{sidewaystable}
	\caption{\label{brresult}
The calculated branching fractions and up-down asymmetries (in units of \%) along with the ones in the  literature. 
All the branching fractions are normalized by $(\tau (\Xi_{cc}^{++}) , \tau(\Xi_{cc}^{+} ))= (2.56, 0.45) \times 10^{-13}s$.
For Ref.~\cite{Sharma:2017txj}, 
we quote the results of the flavor-independent pole, and 
the parentheses of (N) and (H) indicate the form factors are calculated by the nonrelativistic quark model and heavy quark effective theory, respectively. For Refs.~\cite{Ke1,Ke2}, (U) and (M) are the results with and without the $\Xi_c-\Xi_c'$ mixing, respectively.  
}
	\label{NewTable3}
	\begin{tabular}{lccc|ccc|ccc|ccccc}
		\hline
		\hline
		&\multicolumn{3}{c}{HB~$\theta_c= \theta_0 $ } &\multicolumn{3}{|c}{Cheng~{\it et al.}~\cite{Cheng:2020wmk}} &\multicolumn{3}{|c}{Gutsche~{\it et al.}~\cite{Gutsche:2018msz}}&\multicolumn{5}{|c}{Sharma \& Dhir ~\cite{Sharma:2017txj}} \\
	& ~${\cal B}~$ ~& ~$\alpha~$~ & ${\cal R}$	& ~${\cal B}~$ ~& ~$\alpha~$~ & ${\cal R}$& ~${\cal B}~$ ~& ~$\alpha~$~ & ${\cal R}$ & ~${\cal B}$~(N) ~& ~${\cal B}$~(H) ~& ~$\alpha$~ (N)& ~$\alpha$~(H) & ${\cal R}$ \\
		\hline
$\Xi_{cc}^{++} \to \Xi^+ _c \pi^+$	&$10.3(24)$  &$-30$&\multirow{2}{*}{$0.87^{+0.17}_{-0.11} $} & $0.69$&$-41$ &\multirow{2}{*}{$6.74$}&
$0.71$&$-57$&\multirow{2}{*}{$4.77$}&6.66&9.30 &$-99$&$-99$&0.82~(N)\\
$\Xi_{cc}^{++} \to \Xi^{\prime +} _c\pi^{ + }$&$ 8.91(68)$ &$ -96$& & $4.65$&$-84$  &&
$3.39$&$-93$&&5.46&7.51&$-78$&$-79$&0.81~(H)\\
\hline
$\Xi_{cc}^{+} \to \Xi^0_c \pi^+$	&$8.12(55)$  &$-52$&\multirow{2}{*}{$0.25$}&$3.84$ &$-31$& \multirow{2}{*}{$0.40$}&&&&0.59&0.95&55&34&0.39~(N)\\
$\Xi_{cc}^{+} \to \Xi^{\prime 0 }_c \pi^+ $&$ 2.05(17)$ &$ 97$& & $1.55$&$-73$ &&&&&1.49&2.12&65&65&0.45~(H)\\
\hline
$\Xi_{cc}^{+} \to \Xi^+_c \pi^0$	&$8.58(104)$  &$-37$&\multirow{2}{*}{$0.23$}&$2.38$&$-25$ &\multirow{2}{*}{$0.07$}&&&&0.50&&&&\multirow{2}{*}{$0.11$}\\
$\Xi_{cc}^{+} \to \Xi^{\prime +}_c \pi^{0}$&$ 1.94(24)$ &$ 52$& & $0.17$&$-3$&&&&&0.054\\
\hline
\hline
		&\multicolumn{3}{c}{SB~$\theta_c= -\theta_0 $} &\multicolumn{3}{|c}{Shi~{\it et al.}~\cite{Shi:2022kfa}} &\multicolumn{3}{|c}{Gerasimov~{\it et al.}~\cite{Gerasimov:2019jwp}}&\multicolumn{5}{|c}{Ke~{\it et al.}~\cite{Ke1,Ke2}} \\
& ~${\cal B}~$ ~& ~$\alpha~$~ & ${\cal R}$	& \multicolumn{2}{c}{${\cal B}$} & ${\cal R}$& \multicolumn{2}{c}{${\cal B}$} & ${\cal R}$ & ~${\cal B}$ ~(U)& ~${\cal B}$ ~(M)& ~$\alpha~$~(U)& ~$\alpha~$~(M) & ${\cal R}$ \\
\hline
$\Xi_{cc}^{++} \to \Xi^+ _c \pi^+$	&$2.24$  &$-93$&\multirow{2}{*}{$1.45$} & \multicolumn{2}{c}{$6.22(194)$} &\multirow{2}{*}{$1.42(78)$}&
 \multicolumn{2}{c}{$7.01$}&\multirow{2}{*}{$0.83$}&3.48(46)&2.14(18) &$-44(1)$ & $9 (7) ~$&0.56(18)~(U)\\
$\Xi_{cc}^{++} \to \Xi^{\prime +} _c\pi^{ + }$&$ 3.25$ &$ -63$& & \multicolumn{2}{c}{$8.55(62)$}  && \multicolumn{2}{c}{$5.85$} &&1.96(24)&3.0(1)& $-98(1)$ & $-99(1)$&1.41(20)~(M)\\
\hline
$\Xi_{cc}^{+} \to \Xi^0_c \pi^+$	&$2.26$  &$31$&\multirow{2}{*}{$1.17$}&\multicolumn{2}{c}{}& \multirow{2}{*}{}& \multicolumn{2}{c}{$1.23$} &\multirow{2}{*}{$0.85$}&0.61(8)&0.38(3) &$-44(1)$&$9 (7) ~$ &0.56(18)~(U)\\
$\Xi_{cc}^{+} \to \Xi^{\prime 0 }_c \pi^+ $&$2.64 $ &$ -99 $& & \multicolumn{2}{c}{} &&\multicolumn{2}{c}{$1.04$} &&0.35(4)& 0.53(2)&$-98(1)$&$-99(1)$&1.41(20)~(M)\\
\hline
$\Xi_{cc}^{+} \to \Xi^+_c \pi^0$	&$2.01$  &$-5$&\multirow{2}{*}{$0.25$}&& &&&&&\\
$\Xi_{cc}^{+} \to \Xi^{\prime +}_c \pi^{0}$&$ 0.51$ &$ -65$& & &&&&&&\\
		\hline
		\hline	
	\end{tabular}
\end{sidewaystable}
\clearpage
\newpage
\restoregeometry

\section{Conclusion}\label{5}
We have studied the  $\Xi_c -\Xi_c'$ mixing effects in  $\Xi_{cc} \to \Xi_c \pi$ with the soft-meson limit.   The bag model has been employed for the baryon matrix elements with and without removing the CMM. 
We have found that the CMM corrections are sizable as found in the heavy-flavor-conserving decays. 
The branching fractions and up-down asymmetries have been calculated and special attentions have been given to ${\cal R}$. 
In particular, 
for $\Xi_{cc} ^{++} \to \Xi_c^+ \pi^+$
we have obtained  that  $( {\cal B} , {\cal R}) = (10.3(24)\%, 0.87^{+0.17}_{-0.11} )$ and $(2.24\%, 1.45)$ 
with and  without removing the CMM, respectively, which are consistent with the current experimental data. 
To test our theory, we recommend the future experiments to examine 
${\cal R}(\Xi_{cc}^{+} \to \Xi_c^{0}\pi^+)$, which have been  computed as  $0.25$ and $1.17$  in the HB and SB approaches, respectively. 
To probe the $W$-exchange contributions, we recommend the measurement on $\alpha(\Xi_{cc}^+ \to \Xi_c^{(\prime)+} \pi ^+)$ as they are negative in the factorization ansatz but  $0.31~(0.52)$ in the SB~(HB) approach.

\begin{acknowledgments}
We would like to thank Hai-Yang Cheng for the useful discussions. 
This work is supported in part by the National Key Research and Development Program of China under Grant No. 2020YFC2201501 and  the National Natural Science Foundation of China (NSFC) under Grant No. 12147103.
\end{acknowledgments}


\begin{thebibliography}{9}
		
\bibitem{Belle:2021crz}
Y.~B.~Li \textit{et al.} [Belle],
%``Measurements of the branching fractions of the semileptonic decays $\Xi_{c}^{0} \to \Xi^{-} \ell^{+} \nu_{\ell}$ and the asymmetry parameter of $\Xi_{c}^{0} \to \Xi^{-} \pi^{+}$,''
Phys. Rev. Lett. \textbf{127},  121803 (2021).




\bibitem{ALICE:2021bli}
S.~Acharya \textit{et al.} [ALICE],
%``Measurement of the Cross Sections of $\Xi^0_{c}$ and $\Xi^+_{c}$ Baryons and of the Branching-Fraction Ratio BR($\Xi^0_{c} \rightarrow \Xi^-{e}^+\nu_{ e}$)/BR($\Xi^0_{c} \rightarrow \Xi^-\pi^+$) in pp collisions at 13 TeV,''
Phys. Rev. Lett. \textbf{127},  272001 (2021).



\bibitem{SU(3)1}
X.~G.~He, F.~Huang, W.~Wang and Z.~P.~Xing,
%``SU(3) symmetry and its breaking effects in semileptonic heavy baryon decays,''
Phys. Lett. B \textbf{823}, 136765 (2021).

\bibitem{Geng:2022yxb}
C.~Q.~Geng, X.~N.~Jin and C.~W.~Liu,
%``Resolving puzzle in $\Xi_c^0\to \Xi^-e^+\nu_e$ with $\Xi_c-\Xi_c'$ mixing,''
arXiv:2210.07211 [hep-ph].




\bibitem{Jenkins:1996rr}
E.~E.~Jenkins,
%``Heavy baryon masses in the 1/m(Q) and 1/N(c) expansions,''
Phys. Rev. D \textbf{54}, 4515 (1996);
E.~E.~Jenkins,
%``Update of heavy baryon mass predictions,''
Phys. Rev. D \textbf{55}, 10 (1997);
E.~E.~Jenkins,
%``Model-Independent Bottom Baryon Mass Predictions in the 1/N($c$) Expansion,''
Phys. Rev. D \textbf{77}, 034012 (2008).


\bibitem{De0}
M.~Sumihama \textit{et al.} [Belle],
%``Observation of $\Xi(1620)^0$ and evidence for $\Xi(1690)^0$ in $\Xi_c^+ \rightarrow \Xi^-\pi^+\pi^+$ decays,''
Phys. Rev. Lett. \textbf{122}, 072501 (2019).



\bibitem{De1}
J.~G.~Korner and M.~Kramer,
%``Exclusive nonleptonic charm baryon decays,''
Z. Phys. C \textbf{55}, 659 (1992).

\bibitem{De2}
Q.~P.~Xu and A.~N.~Kamal,
%``The Nonleptonic charmed baryon decays: B(c) ---\ensuremath{>} B (3/2+, decuplet) + P(0-) or V(1-),''
Phys. Rev. D \textbf{46}, 3836 (1992).

\bibitem{De3}
K.~K.~Sharma and R.~C.~Verma,
%``SU(3) flavor analysis of two-body weak decays of charmed baryons,''
Phys. Rev. D \textbf{55}, 7067 (1997).

\bibitem{De4}
C.~Q.~Geng, C.~W.~Liu, T.~H.~Tsai and Y.~Yu,
%``Charmed Baryon Weak Decays with Decuplet Baryon and SU(3) Flavor Symmetry,''
Phys. Rev. D \textbf{99},  114022 (2019).



\bibitem{Ratio}
R.~Aaij \textit{et al.} [LHCb],
%``Observation of the doubly charmed baryon decay $ {\Xi}_{cc}^{++}\to {\Xi}_c^{\prime +}{\pi}^{+} $,''
JHEP \textbf{05}, 038 (2022).


\bibitem{Wang:2017mqp}
W.~Wang, F.~S.~Yu and Z.~X.~Zhao,
%``Weak decays of doubly heavy baryons: the $1/2\rightarrow 1/2$ case,''
Eur. Phys. J. C \textbf{77},  781 (2017).


\bibitem{Sharma:2017txj}
N.~Sharma and R.~Dhir,
%``Estimates of W-exchange contributions to $\Xi_{cc}$ decays,''
Phys. Rev. D \textbf{96},  113006 (2017). 


\bibitem{Gerasimov:2019jwp}
A.~S.~Gerasimov and A.~V.~Luchinsky,
%``Weak decays of doubly heavy baryons: Decays to a system of $\pi$ mesons,''
Phys. Rev. D \textbf{100},  073015 (2019).

\bibitem{Gutsche:2018msz}
T.~Gutsche, M.~A.~Ivanov, J.~G.~K\"orner, V.~E.~Lyubovitskij and Z.~Tyulemissov,
%``Ab initio three-loop calculation of the $W$-exchange contribution to nonleptonic decays of double charm baryons,''
Phys. Rev. D \textbf{99},  056013 (2019).



\bibitem{Ke1}
H.~W.~Ke, F.~Lu, X.~H.~Liu and X.~Q.~Li,
%``Study on $\Xi_{cc}\to\Xi_c$ and $\Xi_{cc}\to\Xi'_c$ weak decays in the light-front quark model,''
Eur. Phys. J. C \textbf{80},  140 (2020). 


\bibitem{Cheng:2020wmk}
H.~Y.~Cheng, G.~Meng, F.~Xu and J.~Zou,
%``Two-body weak decays of doubly charmed baryons,''
Phys. Rev. D \textbf{101}, 034034 (2020).



\bibitem{pdg}
R.~L.~Workman \textit{et al.} [Particle Data Group],
%``Review of Particle Physics,''
PTEP \textbf{2022}, 083C01 (2022).


\bibitem{LHCb:2018pcs}
R.~Aaij \textit{et al.} [LHCb],
%``First Observation of the Doubly Charmed Baryon Decay $\Xi_{cc}^{++}\rightarrow \Xi_{c}^{+}\pi^{+}$,''
Phys. Rev. Lett. \textbf{121},  162002 (2018).

\bibitem{Belle:2019bgi}
Y.~B.~Li \textit{et al.} [Belle],
%``First measurements of absolute branching fractions of the $\Xi_c^+$ baryon at Belle,''
Phys. Rev. D \textbf{100},  031101 (2019);
R.~Aaij \textit{et al.} [LHCb],
%``First branching fraction measurement of the suppressed decay $\Xi_c^0\to \pi^-\Lambda_c^+$,''
Phys. Rev. D \textbf{102}, 071101 (2020).


\bibitem{Gutsche:2019iac}
T.~Gutsche, M.~A.~Ivanov, J.~G.~K\"orner, V.~E.~Lyubovitskij and Z.~Tyulemissov,
%``Analysis of the semileptonic and nonleptonic two-body decays of the double heavy charm baryon states $\Xi_{cc}^{++},\,\Xi_{cc}^{+}$ and $\Omega_{cc}^+$,''
Phys. Rev. D \textbf{100},  114037 (2019).




\bibitem{Lcthe}
H.~Y.~Cheng, X.~W.~Kang and F.~Xu,
%``Singly Cabibbo-suppressed hadronic decays of $\Lambda_c^+$,''
Phys. Rev. D \textbf{97},  074028 (2018);
J.~Zou, F.~Xu, G.~Meng and H.~Y.~Cheng,
%``Two-body hadronic weak decays of antitriplet charmed baryons,''
Phys. Rev. D \textbf{101},  014011 (2020). 




\bibitem{HFCLHCb}
R.~Aaij \textit{et al.} [LHCb],
%``Evidence for the strangeness-changing weak decay $\Xi_b^-\to\Lambda_b^0\pi^-$,''
Phys. Rev. Lett. \textbf{115},  241801 (2015).
\bibitem{HFCExp}
R.~Aaij \textit{et al.} [LHCb],
%``First branching fraction measurement of the suppressed decay $\Xi_c^0\to \pi^-\Lambda_c^+$,''
Phys. Rev. D \textbf{102}, 071101 (2020).
\bibitem{HFCBelle}
[Belle],
%``Measurement of the branching fraction of $\Xi_{c}^{0}\to \Lambda_{c}^{+}\pi^{-}$ at Belle,''
arXiv:2206.08527 [hep-ex].
\bibitem{HFC}
H.~Y.~Cheng, C.~W.~Liu and F.~Xu,
%``Heavy-flavor-conserving hadronic weak decays of charmed and bottom baryons: An update,''
Phys. Rev. D \textbf{106},  093005 (2022). 


\bibitem{Groote:2021pxt}
S.~Groote and J.~G.~K\"orner,
%``Topological tensor invariants and the current algebra approach: analysis of 196 nonleptonic two-body decays of single and double charm baryons \textendash{} a review,''
Eur. Phys. J. C \textbf{82},  297 (2022).




\bibitem{Bag1}
A.~Chodos, R.~L.~Jaffe, K.~Johnson, C.~B.~Thorn and V.~F.~Weisskopf,
%``A New Extended Model of Hadrons,''
Phys. Rev. D \textbf{9}, 3471 (1974).

\bibitem{Bag2}
A.~Chodos, R.~L.~Jaffe, K.~Johnson and C.~B.~Thorn,
%``Baryon Structure in the Bag Theory,''
Phys. Rev. D \textbf{10}, 2599 (1974).

\bibitem{Bag3}
T.~A.~DeGrand, R.~L.~Jaffe, K.~Johnson and J.~E.~Kiskis,
%``Masses and Other Parameters of the Light Hadrons,''
Phys. Rev. D \textbf{12}, 2060 (1975).



\bibitem{Liu:2022pdk}
C.~Q.~Geng, C.~W.~Liu and T.~H.~Tsai,
%``Nonleptonic two-body weak decays of $\Lambda_b$ in modified MIT bag model,''
Phys. Rev. D \textbf{102}, 034033 (2020);
C.~W.~Liu and C.~Q.~Geng,
%``Center of mass motion in bag model,''
arXiv:2205.08158 [hep-ph]. 
		
	

\bibitem{Geng:2022uyy}
C.~Q.~Geng, C.~W.~Liu, A.~Zhou and X.~Yu,
%``Semileptonic decays of doubly charmed baryons with bag model,''
arXiv:2211.04372 [hep-ph].


\bibitem{bag}
T.~A.~DeGrand, R.~L.~Jaffe, K.~Johnson and J.~E.~Kiskis,
%``Masses and Other Parameters of the Light Hadrons,''
Phys. Rev. D \textbf{12}, 2060 (1975).


\bibitem{Korner:1970xq}
J.~G.~Korner,
%``Octet behaviour of single-particle matrix elements $ < B'|H(W)|B > $ and $< M'|H(W)|M>$ using a weak current current quark Hamiltonian,''
Nucl. Phys. B \textbf{25}, 282 (1971);
J.~C.~Pati and C.~H.~Woo,
%``$\Delta I$ = 1/2 rule with fermion quarks,''
Phys. Rev. D \textbf{3}, 2920 (1971).



\bibitem{EXPLifetime}
R.~Aaij \textit{et al.} [LHCb],
%``Measurement of the Lifetime of the Doubly Charmed Baryon $\Xi_{cc}^{++}$,''
Phys. Rev. Lett. \textbf{121},  052002 (2018).

\bibitem{LifeTH}
H.~Y.~Cheng and Y.~L.~Shi,
%``Lifetimes of Doubly Charmed Baryons,''
Phys. Rev. D \textbf{98},  113005 (2018).


\bibitem{Shi:2022kfa}
Y.~J.~Shi, Z.~X.~Zhao, Y.~Xing and Ulf-G.~Mei\ss{}ner,
%``W-exchange contribution to the decays \ensuremath{\Xi}cc++\textrightarrow{}\ensuremath{\Xi}c+(')\ensuremath{\pi}+ using light-cone sum rules,''
Phys. Rev. D \textbf{106},  034004 (2022).


\bibitem{Ke2}
H.~W.~Ke and X.~Q.~Li,
%``Revisiting the transition \ensuremath{\Xi}cc++\textrightarrow{}\ensuremath{\Xi}c(')+ to understand the data from LHCb,''
Phys. Rev. D \textbf{105},  096011 (2022).

		
	\end{thebibliography}
\end{document}